\input{aipcheck}

\documentclass[
    ,final            % use final for the camera ready runs
  ]
  {aipproc}

\layoutstyle{8x11single}

\newboolean{uprightparticles}
\setboolean{uprightparticles}{false} %Set to true to get roman particle symbols
\usepackage{amssymb}
\usepackage{amsfonts}
\usepackage{upgreek}
\usepackage{xspace}
\usepackage{lhcb-symbols-def}

\begin{document}

\title{Search for \CP violation in the $\Bs-\Bsb$ system}


\author{Till Moritz Karbach\footnote{on behalf of the LHCb collaboration}}{
  address={TU Dortmund, Germany}
}

\begin{abstract}
We present studies from the LHCb experiment leading to the measurement
of the weak phase $\phi_s$. At first, flavor tagging is established by
measuring the \Bs oscillation frequency \dms. Then, flavor tagging is used
to perform a measurement of the well known CKM angle $\sin2\beta$ in $\Bd\to\jpsi\KS$,
before we constrain $\phi_s$ through an amplitude analysis of $\Bs\to\jpsi\phi$
decays. These studies use about 35\invpb of data taken in 2010. In addition,
we present the measurement of $\BR(\Bp\to\jpsi\pip)/\BR(\Bp\to\jpsi\Kp)$ and
the first observation of $\Bs\to\jpsi f_2'(1525)$.
\end{abstract}

\maketitle

%%%%%%%%%%%%%%%%%%%%%%%%%%%%%%%%%%%%%%%%%%%%
%% MAINMATTER
%%%%%%%%%%%%%%%%%%%%%%%%%%%%%%%%%%%%%%%%%%%%

\section{Introduction}

Decays of neutral $B$ mesons provide a unique laboratory to study \CP violation originating
from a weak phase in the CKM matrix. The relative phase between two amplitudes, direct decay and
decay after mixing, gives rise to time-dependent \CP violation. 
The decay $\Bs\to\jpsi\phi$ is considered the golden mode for measuring this
type of \CP violation in the $\Bs$ system. In the Standard Model, the \CP violating phase
in this decay is predicted to be $\phi_s \approx -2\beta_s$, where $\beta_s = \arg(-V_{ts}V^*_{tb}/V_{cs}V^*_{cb})$.
Indirect measurements show $2\beta_s$ is small, $2\beta_s = (0.0363 \pm 0.0017)\,\rm rad$~\cite{indirectphi}.
But new contributions to $\Bs$--$\Bsb$ mixing may alter the expected value of $\phi_s$~\cite{newphys1, newphys2}.
Previous constraints on $\phi_s$ at the 0.5 rad level have been reported by the 
Tevatron experiments CDF~\cite{CDF} and D0~\cite{D0}. The precise determination 
of $\phi_s$ is one of the key goals of the LHCb experiment~\cite{LHCb}.
In this letter we present a series of measurements leading the way to
a measurement of $\phi_s$.

%%%%%%%%%%%%%%%%%%%%

\section{Constraints on $\phi_s$}
%\section{Measurement of $\Delta m_s$}

\begin{figure}
  \includegraphics[height=.225\textheight]{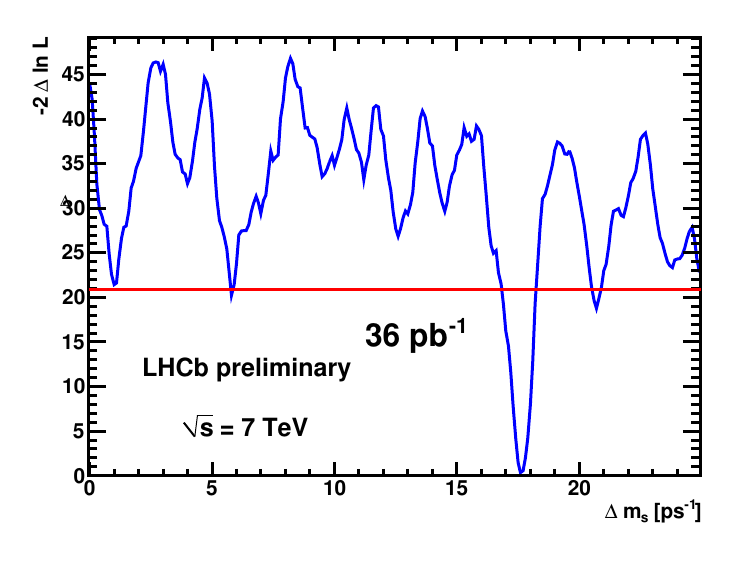}
  \includegraphics[height=.225\textheight]{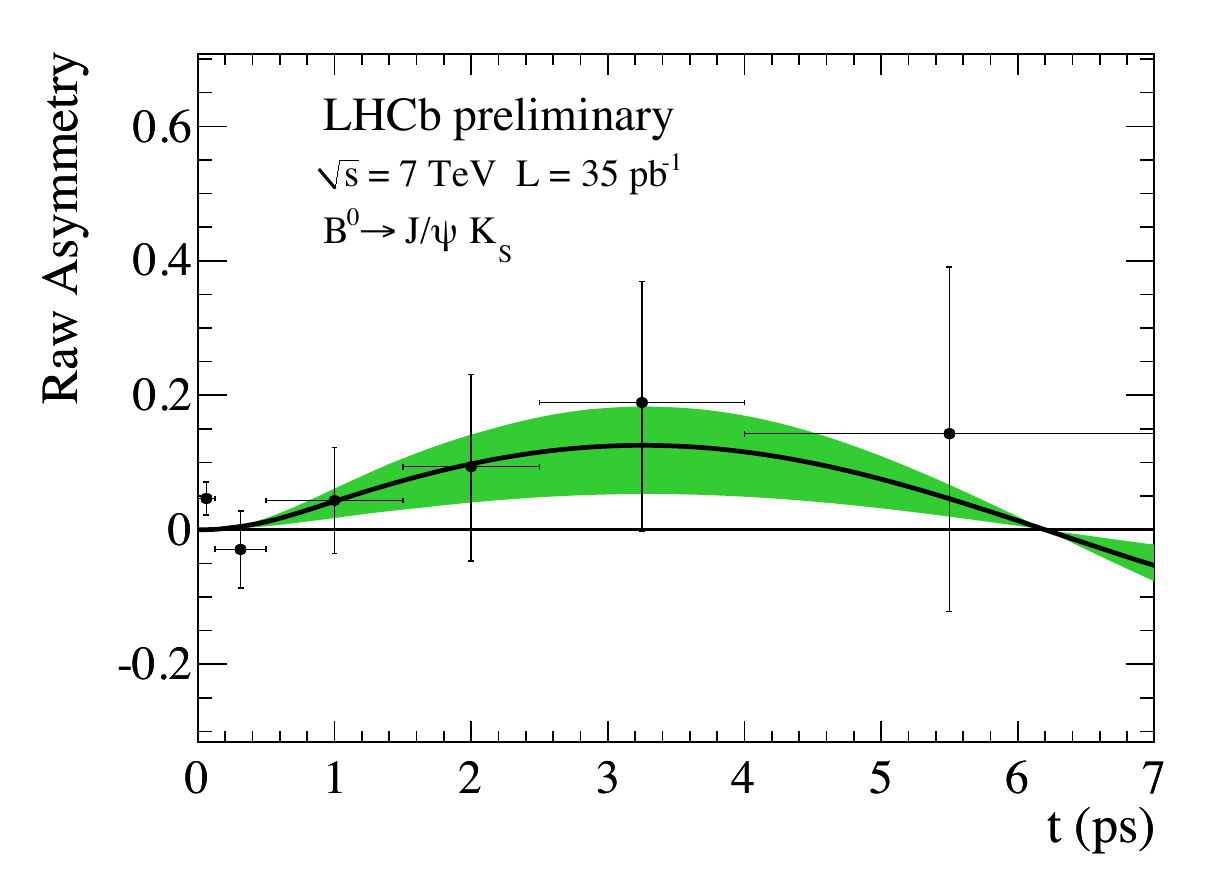}
  \label{fig:deltamsSin2beta}
  \caption{Left: Likelihood scan for $\Delta m_s$ in the range from $[0,25] \ps^{-1}$.
  The horizontal line corresponds to the likelihood value expected for infinitely fast oscillations.
  Right: Time dependent raw \CP asymmetry in $\Bd\to\jpsi\KS$ with the fit projection (signal and background) overlaid. The green band corresponds to the one standard deviation statistical error.}
\end{figure}

We first present~\cite{deltams} a measurement of the mixing frequency $\Delta m_s$ of the
\Bs system, using about 1350 \Bs signal candidates reconstructed in $36\invpb$ collected in 2010
through their decays $\Bs\to D_s^-\pi$ and $\Bs\to D_s^- 3\pi$, where $D_s^-\to\Kp\Km\pim$.
This measurement demonstrates that the achieved decay time resolution ($44\fs$ in $D_s^-\pi$, 
$36\fs$ in $D_s^- 3\pi$) is sufficient to resolve the fast oscillations in \Bs mixing. It
also establishes the flavor tagging algorithms required to identify the \Bs flavor at
production time. The effective tagging efficiency is $\varepsilon_{\rm eff} = 3.8\pm2.1\stat \%$.
Figure~\ref{fig:deltamsSin2beta} shows the result of an amplitude scan, converging to
$\Delta m_s = 17.63 \pm 0.11\stat \pm 0.04\syst \ps^{-1}$, which is 
compatible with the Tevatron measurements and of similar precision.

%%%%%%%%%%%%%%%%%%%%

%\section{Measurement of $\sin2\beta$}

We then~\cite{sin2beta} make use of the flavor tagging algorithms to repeat
the measurement of the CKM angle $\sin2\beta$, which was determined by the
B factories to amazing precision: $\sin2\beta = 0.673 \pm 0.023$~\cite{sin2betaHfag}.
In $35\invpb$ we find about 280 tagged $\Bd\to\jpsi\KS$ decays, considered the
golden channel for this measurement. A maximum likelihood fit to the decay time
distribution and the \Bd invariant mass, simultaneously fitting tagged and untagged
samples, reports $\sin2\beta \simeq S = 0.53^{+0.28}_{-0.29}\stat \pm 0.05\syst$.
The systematic uncertainty is dominated by that on the flavor tagging. 
Although not yet competitive with the result of the B-factories a very precise 
measurement will be possible with the data that LHCb will collect over the 
coming few years.
Figure~\ref{fig:deltamsSin2beta}
shows the time dependent raw asymmetry of \Bd and \Bdb decaying into $\jpsi\KS$,
the amplitude of which is proportional to $S$.

%%%%%%%%%%%%%%%%%%%%

%\section{Constraints on $\phi_s$}

\begin{figure}
  \includegraphics[height=.225\textheight]{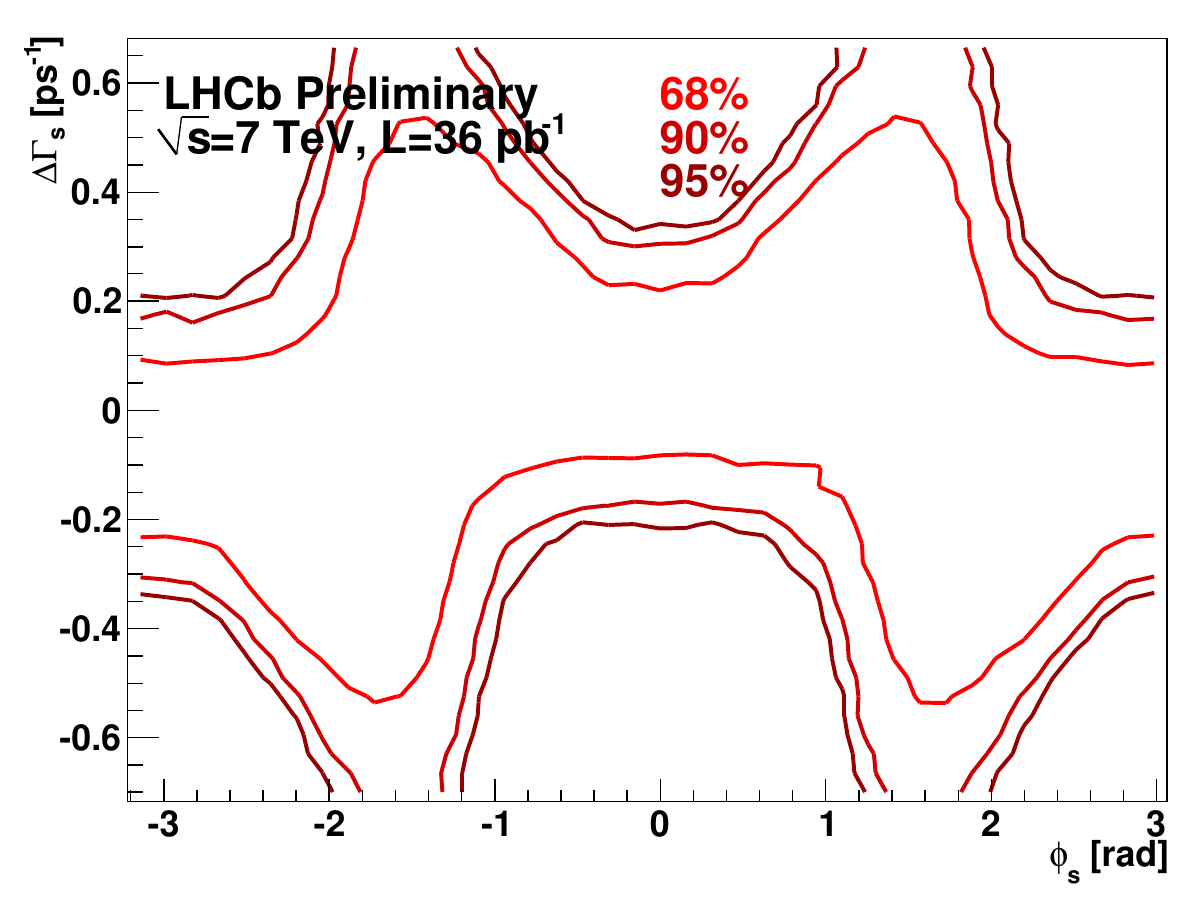}
  \includegraphics[height=.225\textheight]{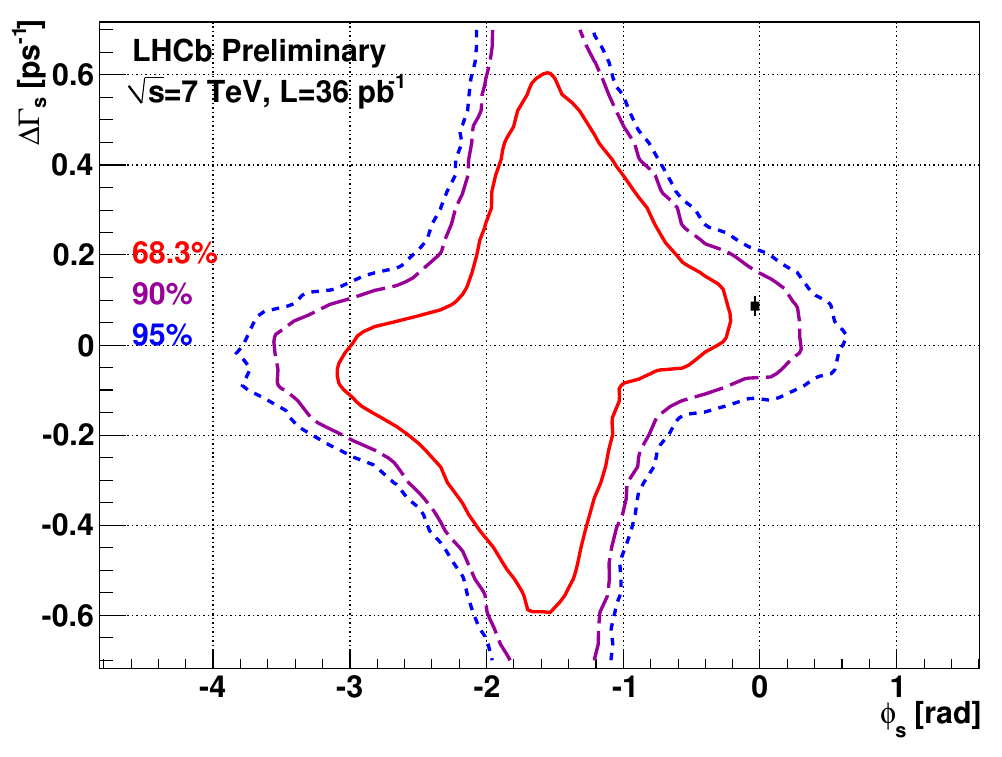}
  \label{fig:phis}
  \caption{Feldman-Cousins confidence regions in the $\Delta\Gamma_s-\phi_s$ plane.
  The Standard Model estimate is at the black square. Both untagged (left) and
  tagged (right) analyses of $\Bs\to\jpsi\phi$ decays are shown.}
\end{figure}

Finally, we perform~\cite{phisuntagged} an untagged angular analysis 
of $\Bd\to\jpsi\Kstarz$ and $\Bs\to\jpsi\phi$ decays. This gives access 
to the decay amplitudes for both final states, as well as the lifetime 
and lifetime difference $\Delta\Gamma_s$ for $\Bs\to\jpsi\phi$. Due to 
the forward geometry of the LHCb detector, the reconstruction efficiency 
for these decays is a non-trivial function of the decay angles.
Then we add~\cite{phistagged} the flavor tagging information to the
measurement. In the $36\invpb$ of 2010 we find approximately 760
$\Bs\to\jpsi\phi$ events. We extract constraints on $\phi_s$ through a maximum
likelihood fit to the decay time distribution, the \Bs invariant mass,
and three decay angles. The data sample is too small to allow for meaningful
point estimates. Instead we perform a Feldman-Cousins analysis which
gives contours in the $\Delta\Gamma_s-\phi_s$ plane (Figure~\ref{fig:phis}) that have frequentist
coverage. We observe a deviation from the Standard Model of 1.2$\sigma$,
and constrain $\phi_s \in [-2.7, -0.5]\,\rm rad$ at the $68.3$\% confidence level.

%%%%%%%%%%%%%%%%%%%%

\section{Analysis of $\Bs\to\jpsi (KK, \pi\pi)$}

Following our recent first observation of the $\Bs\to\jpsi f_0$ decay~\cite{f0firstobs},
we update~\cite{f0f2} the analysis of the $\Bs\to\jpsi\pi\pi$ final state with a larger dataset of
$162\invpb$, coming mostly from the first period of the 2011 run.
We also show that the $f_0$ resonance is consistent with being purely $S$-wave,
making $\jpsi f_0$ a pure \CP odd eigenstate. This will allow for a measurement of $\phi_s$ in 
$\Bs\to\jpsi f_0$ decays without the need of an amplitude analysis. We measure the ratio
of rates in a $\pm90\mevcc$ mass window around the $f_0(980)$, $R^{f_0}_{\rm effective} 
= \BR(\Bs\to\jpsi f_0, f_0\to\pi\pi)/\BR(\Bs\to\jpsi\phi, \phi\to KK)$,
to be $R^{f_0}_{\rm effective} = (21.7 \pm 1.1\stat \pm 0.7\syst)\%$.

\begin{figure}
  \includegraphics[height=.225\textheight]{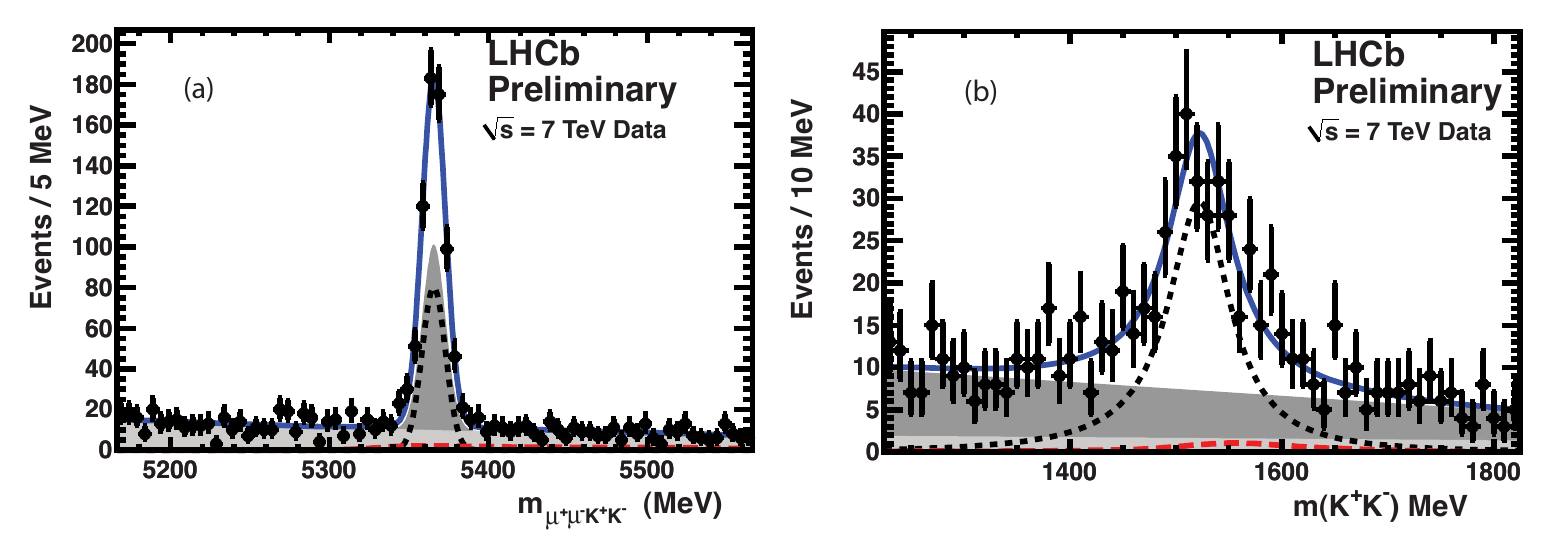}
  \label{fig:f2}
  \caption{The $\jpsi\Kp\Km$ (a) and the $\Kp\Km$ (b) invariant mass spectra showing peaks
  of the \Bs and, for the first time in this mode, the $f_2'(1525)$ (short dashed). Shown are the
  combinatorial background (light grey), non-resonant $\jpsi\Kp\Km$ (dark grey),
  $\Bd\to\jpsi\Km\pip$ background (long-dashed).}
\end{figure}

We extend our analysis to the $\Bs\to\jpsi\Kp\Km$ final state. In the $\Kp\Km$
invariant mass spectrum we for the first time observe, in addition to the $\phi(1020)$ component, 
a structure that we identify as the spin-2 $f_2'(1525)$. An angular analysis confirms,
that the data are consistent with the spin-2, and inconsistent with the spin-0  
hypothesis. The $\Bs\to\jpsi f_2'$ mode can also be used to measure $\phi_s$, although here a 
transversity analysis would be required as in $\jpsi\phi$. It is also possible that 
this mode could be used to resolve ambiguities in $\phi_s$ if the interference with 
non-resonant $\jpsi\Kp\Km$ is significant. Figure~\ref{fig:f2} shows both the $\jpsi\Kp\Km$
and the $\Kp\Km$ invariant mass spectra for this first observation.
We measure $R^{f_2'}_{\rm effective} = (19.4 \pm 1.8\stat \pm 1.1\syst)\%$
in a $\pm125\mevcc$ mass window around the $f_2'$.

%%%%%%%%%%%%%%%%%%%%

\section{Measurement of $\BR(\Bp\to\jpsi\pip)/\BR(\Bp\to\jpsi\Kp)$}

We also analyze~\cite{br} the $\Bp\to\jpsi\pip$ and $\Bp\to\jpsi\Kp$ decay channels,
of which the latter plays an important role in the calibration of the flavor tagging
algorithms. In $37\invpb$ we measure the ratio of their branching fractions to be
$\BR(\Bp\to\jpsi\pip)/\BR(\Bp\to\jpsi\Kp)=(3.94 \pm 0.39\stat \pm 0.17\syst) \times 10^{-2}$.
This result has a precision comparable to the present world average~\cite{brwa}, but
is lower by $2.2\sigma$.

%\begin{figure}
%  \includegraphics[height=.225\textheight]{br}
%  \caption{Picture to fixed height}
%\end{figure}

%%%%%%%%%%%%%%%%%%%%%%%%%%%%%%%%%%%%%%%%%%%%%%%%
%% BACKMATTER
%%%%%%%%%%%%%%%%%%%%%%%%%%%%%%%%%%%%%%%%%%%%%%%%

\bibliographystyle{aipproc}   % if natbib is available

\end{document}